# COMPARATIVE ANALYSIS OF POX AND RYU SDN CONTROLLERS IN SCALABLE NETWORKS

Chandimal Jayawardena, Jay Chen, Amay Bhalla, and Lin Bui

STEM, University of South Australia, Adelaide, SA, 5095, Australia.

## ABSTRACT

*This paper explores the Quality of Service (QoS) performance of two widely used Software-Defined Networking (SDN) controllers, POX and Ryu, using Mininet for network simulation. SDN, a transformative approach to network architecture, separates the control and data planes, enabling centralized management, improved agility, and cost-effective solutions. The study evaluates key QoS parameters, including throughput, delay, and jitter, to understand the capabilities and limitations of the POX and Ryu controllers in handling traffic under diverse network topologies.*

*The research employs a systematic methodology involving the design of custom network topologies, implementation of OpenFlow rules, and analysis of controller behavior under simulated conditions. Results reveal that while POX offers simplicity and ease of use, making it suitable for smaller-scale applications and experimentation, Ryu provides superior scalability and adaptability for more complex network environments.*

*The findings highlight the strengths and challenges of each controller, providing valuable insights for organizations seeking to optimize SDN deployment. This study contributes to the growing body of knowledge on SDN technologies and their role in building scalable, efficient, and resilient network infrastructures.*

## 1. INTRODUCTION

In today's rapidly advancing technological landscape, networks have become increasingly complex. Traditional network architectures that require significant resources for updates or even a complete overhaul to support expanding business needs are no longer viable options. Businesses now seek networks that offer scalability, flexibility, and redundancy to support future growth. These complex networks involve numerous switches, routers, and end devices. As a result, making changes to the network requires manually adjusting configurations on all devices, which is often unfeasible for network administrators.

To address this challenge, the growing trend of automated network management has led to the introduction of Software-Defined Networking (SDN) [1]. SDN has revolutionized network management by decoupling the control and data planes of network devices. In traditional network architectures, the data and control planes are integrated within network devices, necessitating constant communication and updates about network traffic. In contrast, SDN separates these planes, with the control plane typically centralized and managed by a software-based controller, while the data plane consists of network devices that handle traffic based on the controller's instructions [2]. This architectural separation enables SDN to offer several advantages in complex

 

Networks, including centralized management, reduced operational costs, simplified network configuration, increased agility and flexibility, and improved scalability [3]. A simplified representation of SDN architecture is shown in Fig. 1. At the top of the architecture, applications interact with the SDN controller via the northbound interface, which allows applications to define network policies and behaviors through APIs. The SDN controller, serving as the central intelligence of the network, manages traffic flows and network resources by communicating with underlying network devices through the southbound interface. This interface uses protocols like OpenFlow to relay instructions to switches, routers, and other devices. This architecture enhances network flexibility, programmability, and efficiency by enabling dynamic adjustments to network conditions based on application needs.

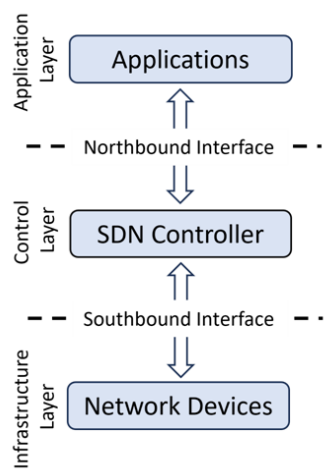

Figure1. SDNArchitecture

Organizations can greatly benefit from Software-Defined Networking (SDN) by gaining better control over their network infrastructure, improving efficiency, and reducing costs. SDN centralizes network management, making it easier to configure, monitor, and maintain networks, especially as they grow [4]. By separating the control plane from the data plane, SDN enables more flexible and dynamic network management, allowing businesses to respond to changing demands and optimize traffic flow quickly. It also lowers operational costs by automating network provisioning, enhancing resource allocation, and reducing manual configuration errors. SDN's ability to integrate multiple vendors and technologies within a unified framework improves interoperability and reduces dependency on a single vendor, offering organizations a more adaptable and cost-effective solution. Moreover, SDN enhances security by providing granular control over network traffic and enabling a rapid response to potential threats. Ultimately, SDN helps organizations build agile, scalable, and efficient networks that support business growth and innovation [5].

The SDN controller is the central component of any SDN architecture, as it is responsible for managing and controlling the network's behavior by communicating with the data plane devices such as routers and switches and making decisions regarding traffic flow and network policies [6]. Some of the most widely used SDN controllers include NOX [7], POX [8], Floodlight [9], OpenDaylight [10], ONOS [11], and Ryu [12]. In addition to these, many other controllers exist in literature. [13] presents a comprehensive list of controllers with their features.

The choice of an SDN controller is highly dependent on the specific priorities and requirements of the organisation. For example, large enterprises with complex network topologies might prefer controllers like OpenDaylight or Floodlight due to their scalability and advanced features, while smaller organizations or those with less complex network needs might choose a more lightweight controller like Ryu or POX. Moreover, the controller must align with the organisation's connectivity needs, such as support for various network protocols, compatibility with hardware devices, and the ability to manage high levels of traffic. Factors like security, ease of integration, and the availability of developer tools for customisation also play crucial roles in controller selection. Ultimately, selecting the right SDN controller ensures that the network can evolve with the organisation's changing demands, enhancing both performance and agility. In this research, Quality of Service (QoS) comparison is performed for two selected SDN controllers: POX and Ryu. QoS comparison of SDN controllers is important as it directly impacts the performance, reliability, and user experience of a network. Poor QoS can lead to packet loss, jitter, and delay, which negatively impact the overall user experience. Moreover, different applications have different QoS requirements based on their latency and bandwidth requirements. Therefore, a thorough understanding of QoS parameters of SDN controllers is crucial for selecting the right SDN controller for a network.

The rest of the paper is structured as follows. Section 2 presents the background to the research presented in this paper. In Section 3, the methodology is presented. Results and analysis are presented in Section 4. Section 5 presents the conclusion.

## 2. BACKGROUND

QoS in computer networks is a broad concept that can be understood from two perspectives: the user perspective and the network perspective. From the user perspective, QoS refers to the quality of service experienced for a subscribed service, typically evaluated using parameters such as latency, jitter, and packet loss. This perspective is crucial for assessing network performance and user satisfaction. From the network perspective, QoS is the capability of the network to deliver the level of service perceived by the end user. Achieving this requires two key capabilities: differentiating between various types of traffic—such as data, video, and audio—and applying appropriate policies to ensure optimal performance [14]. In the context of this research, QoS is primarily considered from the end-user perspective.

QoS is particularly important in SDN environments, where diverse applications—ranging from real-time video streaming to large-scale data transfers—have varying performance requirements. The SDN controller plays a critical role in traffic management and resource allocation, ensuring that network policies align with QoS demands [15]. Effective controller performance is essential to maintaining high levels of network efficiency and user satisfaction.

Several SDN controllers have been developed to address different network requirements, with POX and Ryu being two widely studied options. These lightweight controllers are well-suited for academic research and experimental setups.

POX, one of the earliest SDN controllers, is designed for simplicity and rapid prototyping [8]. Written in Python and available as open-source software, POX provides an accessible platform for educational environments and small-scale networks. Derived from its predecessor NOX [7], POX facilitates communication with network devices through the OpenFlow protocol, currently supporting version 1.0, with potential upgrades to version 1.3 in the future to align with industry standards. Despite its limited scalability and advanced features, POX remains valuable for understanding the core principles of SDN [16].

In contrast, Ryu offers a more versatile and feature-rich platform, supporting multiple OpenFlow versions and other protocols. This flexibility allows developers to customise and integrate new functionalities, making Ryu suitable for complex network scenarios such as traffic engineering, resource optimisation, and policy enforcement [12]. Like POX, Ryu is also developed in Python and is open-source under the Apache 2.0 license [17]. Hosted on GitHub, Ryu is actively maintained by the community, ensuring continuous improvements and accessibility. RYU provides a flexible platform for network traffic management through well-defined application programming interfaces (APIs), allowing network administrators to implement and control policies effectively [2]. Moreover, RYU supports business-oriented functionalities such as user account management and policy enforcement.

Comparing the QoS performance of POX and RYU is significant for several reasons. First, these controllers represent different ends of the spectrum in terms of capabilities and complexity. POX, being lightweight and straightforward, provides insights into baseline QoS performance, while RYU's advanced features and scalability make it suitable for analysing complex network environments. Second, both controllers are well-documented and widely used in research, facilitating comprehensive comparisons. Third, evaluating their QoS performance helps in selecting the appropriate controller based on specific network requirements, particularly for applications with strict latency or bandwidth demands.

QoS evaluation typically involves analysing key parameters such as throughput, latency, jitter, and packet loss, which directly impact network reliability and user experience. For example, real-time applications like video conferencing and online gaming require low latency and minimal jitter, whereas data-intensive tasks such as file transfers prioritise high throughput. Comparing POX and RYU across these parameters provides valuable insights for optimising network performance.

## 2.1. Performance Comparison of POX and Ryu

Several studies have compared the performance of POX and RYU controllers across various parameters. Their comparative evaluation highlights differences in throughput, latency, jitter, and packet loss, with each controller demonstrating strengths in specific contexts.

Studies such as [18] have shown that RYU outperforms POX in terms of throughput and latency in different topologies using the Mininet environment. Similarly, [19] found that RYU offers superior QoS by minimising average packet loss and delay while running the Spanning Tree Protocol (STP). Additional research by [20] and [16] has reinforced these findings, concluding that RYU provides better delay, jitter, bitrate, and overall performance in high-throughput applications. Conversely, some studies favour POX in specific scenarios. [17] evaluated both controllers across jitter, throughput, packet loss, and packet delivery ratio, concluding that POX outperforms RYU. Similarly, [21] observed that POX exhibits lower delay and jitter compared to RYU. In a hybrid network setting involving both normal and SDN switches, [22] found that POX delivered superior throughput, although other performance metrics were not considered.

Some comparative studies have found that the performance of POX and RYU varies depending on network conditions. [23] demonstrated that RYU provides better latency, while POX excels in throughput in Simple-Tree-Based and Fat-Tree-Based networks. Similarly, [24] found that while POX offers better throughput, RYU achieves a higher packet delivery ratio when tested using Dijkstra's shortest path algorithm. According to [25], POX achieves higher throughput, whereas RYU offers lower delay, with both controllers showing similar results in packet loss.

It is important to note that performance variations in these studies may also stem from differences in data collection methodologies, experimental configurations, and network conditions. To ensure the reliability of our analysis, this study adopts a structured and systematic data collection

International Journal of Computer Networks & Communications (IJCNC) Vol.17, No.2, March 2025approach. Multiple experimental runs were conducted under controlled conditions to minimize random fluctuations, and performance metrics were averaged to reduce the impact of outliers. Additionally, standardized network topologies were used to ensure consistency across tests. By carefully designing the methodology to limit measurement inconsistencies, this study provides a more accurate and reproducible comparison of the POX and RYU controllers.

In summary, gaining a comprehensive understanding of the QoS and performance characteristics of the SDN controllers, such as POX and RYU, is essential for making informed decisions when selecting the most appropriate controller for various network environments. Each SDN controller offers distinct features, performance metrics, and capabilities that impact critical aspects such as latency, throughput, scalability, and reliability. This research aims to conduct a thorough evaluation of the POX and Ryu controllers, systematically assessing their strengths and limitations in handling diverse network conditions. By analyzing their effectiveness in meeting QoS requirements, the study seeks to provide valuable insights into their suitability for different applications, ranging from small-scale deployments to large, complex network infrastructures. Further more, the findings of this research will contribute to a better understanding of how these controllers can address the evolving challenges and growing demands of modern networks, facilitating more efficient and optimized SDN implementations

## 3. METHODOLOGY

The QoS evaluation of the selected SDN controllers is divided into three distinct stages to ensure a structured and systematic approach. The first stage involves the design and implementation of the simulation environment. This was achieved using the Mininet [26] network simulator. In the second stage, controller rules are implemented in Python for both POX and RYU controllers and applied across all five topologies to enable effective network management. Finally, the third stage focuses on performance evaluation, where the controllers are assessed based on three critical parameters: throughput, delay (round-trip time), and jitter.

Throughput, delay, and jitter were chosen as QoS parameters for comparison as these parameters are considered key QoS parameters. These parameters collectively define the performance and reliability of a network. Throughput measures the amount of data successfully transmitted over the network in a given period. Throughput directly affects the network's ability to handle traffic efficiently. Delay refers to the time taken for a packet to travel from the source to the destination. Lower delay is essential for real-time applications. Jitter represents the variability in packet delay over time. Low jitter or consistent delay is critical for maintaining the quality of real-time communications.

### 3.1. Simulation Environment

As mentioned above, Mininet was used as the simulation environment for this research. Mininet is a powerful tool for simulating OpenFlow-based networks. As an open-source platform, it provides comprehensive support for working with SDN networks. Mininet enables the emulation of an entire OpenFlow network on a single machine by creating a realistic virtual environment [27]. With Mininet, users can generate SDN components, configure them as needed, and integrate them into other networks. It supports interactions between key elements such as hosts, switches, con- trollers, and links. Networks can be customized through Python APIs for greater flexibility or constructed using the Command-Line Interface (CLI) for simpler topologies. Mininet is compatible with various SDN controllers, including POX and RYU controllers, making it a versatile option for researchers [28].

39

The topologies are a fundamental component of this project, designed with varying architectures to enable comprehensive testing. Five distinct network topologies were created using MiniEdit, which is Mininet's graphical user interface. Key Mininet components, such as the SDN controller, OpenFlow switches, and hosts, are integrated into MiniEdit to build and manage these topologies effectively. Fig. 2 shows the five topologies created on Mininet.

Each topology includes an SDN controller that connects to all switches, enabling centralized flow control within the network. The five topologies from A to E were designed to represent networks of increasing complexity. These carefully designed topologies facilitate the evaluation and testing of SDN controller functionalities across diverse network structures. Details of each topology are given in the next section.
.
### 3.2. Controller Rules Implementation

The second stage involves implementing controller rules for POX and Ryu using Python across all five network topologies. Python is the preferred programming language for both controllers due to its simplicity and compatibility.

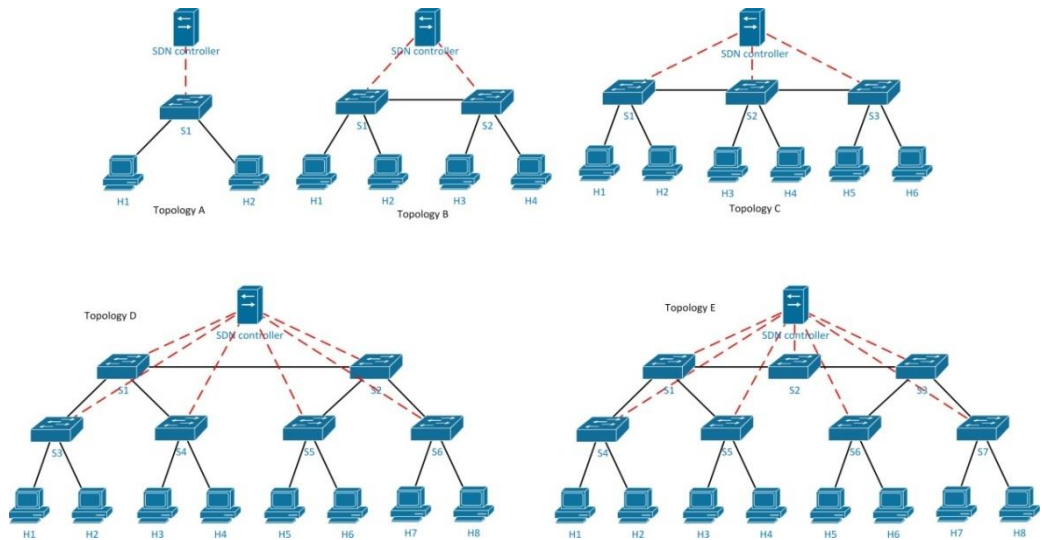

Figure 2 : Topologies Used for Data Collection

The implementation begins by creating Python scripts using Visual Studio Code, an open-source development environment. Each topology is assigned a unique set of controller rules, resulting in a total of 10 Python scripts—five for POX and five for Ryu. These rules are tailored to the specific requirements of each controller, as their parameters and configurations differ.

As mentioned above, Mininet was used to simulate the network. While Mininet excels at creating and visualising network topologies, its switches lack the inherent ability to establish connections with controllers as real switches do. The SDN controller bridges this gap by providing appropriate flow rules that enable the network to function effectively. These flow rules are manually crafted for each topology, ensuring precise communication between switches, hosts, and the controller. Topology-Specific Controller Implementation details are explained below.

- **Topology A**: A basic setup with a single Layer 2 switch connected to two hosts (H1 and H2). The switch is directly connected to the controller, facilitating communication between the hosts via two distinct ports.

- **Topology B**: This topology includes two interconnected Layer 2 switches, each connected to two hosts. The controller links to both switches, enabling traffic flow between the hosts (H1 and H2 to H3 and H4, and vice versa). Each switch features three connections: two to the hosts and one to the other switch.

- **Topology C:** An extension of Topology B, this network adds an additional Layer 2 switch (S2) between S1 and S3. Each of the three switches connects to two hosts and is interlinked. S1 and S3 each have three ports involved, while S2 has four. The controller connects to all three switches to manage flow effectively.

- **Topology D**: This hierarchical or tree network introduces an extra layer of switches. Two Layer 3 switches are each connected to two Layer 2 switches, which in turn connect to two hosts each. The Layer 3 switches are interconnected, enabling inter-switch routing and seamless data packet traversal across the network.

- **Topology E**: Similar to Topology D, this topology adds an extra Layer 3 switch (S2) between S1 and S3, simulating a more complex hierarchical network. The additional Layer 3 switch enhances the network's structure, combining Layer 3 and Layer 2 switches for efficient data flow.

This stage is critical for integrating POX and Ryu controllers with the Mininet environment, enabling comprehensive testing and evaluation of SDN functionalities across diverse network topologies. Each topology's tailored flow rules allow for an in-depth exploration of controller behaviour and network performance.

## 3.3. Testing and Evaluation

The third stage focuses on evaluating the performance of the POX and Ryu controllers across all five network topologies. With the testbed, network topologies, and controller flow rules carefully designed and implemented, this stage is dedicated to assessing the QoS and comparing the performance of the two SDN controllers.

Each topology was tested using both POX and Ryu controllers, and data related to throughput, delay, and jitter were collected. By analysing these metrics for each topology and controller, the testing aimed to identify which controller demonstrates superior performance under different topologies. This comprehensive evaluation provides valuable insights into the capabilities of POX and Ryu, highlighting their strengths and suitability for different network scenarios.

### 3.3.1. Throughput

Measuring network throughput is a critical step in evaluating traffic efficiency, as it provides a clear understanding of the volume of transactions or requests successfully processed within a network. This metric is particularly vital for high-traffic applications, where performance and reliability are paramount. In the context of data transmission, network throughput refers to the amount of data successfully transferred from one point to another within a specified time frame. Throughput is commonly expressed in units such as bits per second (bps), megabits per second (Mbps), or gigabits per second (Gbps) (Burke, 2022).

Throughput testing was conducted for each topology using the 'iperf' a widely utilized tool capable of generating TCP and UDP data streams [29]. This tool can be used to measure the network's throughput, providing insights into its performance and capacity to handle varying traffic loads.

### 3.3.2. Delay

When evaluating network performance, two key types of delay are considered: Propagation Delay and Round-Trip Time (RTT). Propagation Delay refers to the time it takes for a data packet to travel from the source to the destination within a network. In contrast, RTT measures the total time required for a data packet to travel from the source to the destination plus the time it takes to receive an acknowledgement. RTT includes the propagation delay between two points. In this research, RTT was selected as the measure of delay for comparing POX and Ryu. Controllers with lower delays are generally more desirable, as they offer faster response times for transmitting data from the source to the destination and returning information to the source.

RTT is typically measured in milliseconds (ms) using the 'ping' command in a command prompt. To evaluate the performance of the two controllers, a series of 10 pings were sent from one host to another for comparison.

### 3.3.3. Jitter

End devices communicate by exchanging packets of information, which are continuously transmitted over the network. Jitter refers to the variation in packet delay and can impact network performance. The faster the packets travel, the less significant the impact of jitter on network operations. Jitter is typically measured in milliseconds (ms) as the average variation in delay values observed in ping results.

## 4. TEST RESULTS

Throughput, delay, and jitter measurements were conducted between multiple pairs of hosts, including connections such as H1 to H2, H2 to H3, and other relevant host combinations within the network. These measurements were carried out to assess the performance and reliability of data transmission under various topologies. The collected data provides valuable insights into the performance of each controller.

In the following subsections, a detailed presentation of the collected data is provided, along with a comprehensive analysis of the observed patterns. The results are analysed in terms of throughput, which measures the rate of successful data transfer; delay, which indicates the time taken for data to travel from the source to the destination;, and jitter, which captures variations in packet arrival times. Each metric is examined across different host pairs to understand network behaviour under varying network topologies and SDN controllers..

### 4.1. Throughput Comparison

Tables 1 and 2 show the average throughput between different pairs of hosts in each topology. The results in these tables reveal notable differences in average throughput performance between the POX and RYU controllers across various network topologies. For Topology A, RYU demonstrates superior performance, achieving a significantly higher throughput of 70.00 Mbps compared to POX's 55.75 Mbps for H1 to H2. In Topology B, RYU also outperforms POX, with more consistent throughput (56.20 and 56.10 Mbps for H1 to H2 and H2 to H3, respectively) compared to POX's lower values (48.30 and 42.95 Mbps). However, in Topology C, the performance is mixed— while POX achieves higher throughput for H1 to H2 (51.50 Mbps versus RYU's 38.85 Mbps), RYU outperforms POX on H2 to H6 (67.20 Mbps versus 40.35 Mbps). In Topology D, RYU generally provides better performance, particularly on H1 to H8 (71.85 Mbps compared to POX's 56.50 Mbps). Conversely, POX demonstrates superior throughput in Topology E, with consistently higher values across all paths (e.g., 63.05 Mbps for H4 to H5 and 61.15 Mbps for H6 to H8) compared to RYU's relatively lower values (36.70 Mbps for both paths). Overall, RYU excels in Topologies A and D, while POX shows strength in Topology E, highlighting differences in their performance characteristics under varying network configurations.

Fig. 3 shows the overall average throughput for each topology under the POX and RYU controllers. These averages were calculated considering all pairs of hosts under each topology.

Fig. 3 highlights that RYU outperforms POX in Topologies A, B, and D, with notable differences in average throughput. For example, in Topology A, RYU achieves 70 Mbps compared to POX's 55.75 Mbps, a clear improvement. Similarly, in Topology D, RYU leads with 57.92 Mbps, significantly higher than POX's 48.09 Mbps. In Topology B, RYU maintains an advantage with 56.15 Mbps versus POX's 45.63 Mbps.

However, POX demonstrates better performance in Topology E, achieving an average throughput of 55.03 Mbps compared to RYU's much lower 37.21 Mbps. In Topology C, the two controllers exhibit comparable performance, with POX slightly ahead at 51.53 Mbps versus RYU's 50.86 Mbps.

Overall, the results indicate that RYU provides superior throughput in most topologies.

Table 1: Average Throughput Between Hosts (MBits/s) – POX

| Topology | H1 to H2 | H2 to H3 | H4 to H5 | H6 to H8 | H2 to H6 | H1 to H8 |
|---|---|---|---|---|---|---|
| A | 55.75 | - | - | - | - | - |
| B | 48.30 | 42.95 | - | - | - | - |
| C | 51.50 | 49.10 | 65.15 | - | 40.35 | - |
| D | 41.60 | 57.15 | 38.80 | 43.90 | 50.60 | 56.50 |
| E | 51.11 | 45.4 | 63.05 | 46.65 | 62.80 | 61.15 |

Table 2 : Average Throughput Between Hosts (MBits/s) – RYU

| Topology | H1 to H2 | H2 to H3 | H4 to H5 | H6 to H8 | H2 to H6 | H1 to H8 |
|---|---|---|---|---|---|---|
| A | 70.00 | - | - | - | - | - |
| B | 56.20 | 56.10 | - | - | - | - |
| C | 38.85 | 48.35 | 49.05 | - | 67.20 | - |
| D | 43.05 | 63.90 | 68.55 | 58.90 | 41.25 | 71.85 |
| E | 41.35 | 39.05 | 36.70 | 36.70 | 33.40 | 36.05 |

## 4.2. Delay Comparison

Delay was measured between the hosts H1-H2, H2-H3, H4-H5, H6-H8, H2-H6, and H1-H8 (refer to 2 covering all applicable topologies. Tables 3 and 4 show the Round Trip Time (RTT) delay between pairs of hosts under each topology. Values shown in these tables are rounded off to two decimal places to display the tables clearly.

Table 3: Average Delay Between Hosts(ms) – POX

| Top. | Pkt. Size | H1to H2 | H2to H3 | H4to H5 | H6to H8 | H2to H6 | H1to H8 |
|---|---|---|---|---|---|---|---|
| A | 128 | 0.13 | - | - | - | - | - |
|   | 192 | 0.14 | - | - | - | - | - |
|   | 256 | 0.23 | - | - | - | - | - |
| B | 128 | 0.13 | 0.21 | - | - | - | - |
|   | 192 | 0.10 | 0.14 | - | - | - | - |
|   | 256 | 0.11 | 0.13 | - | - | - | - |
| C | 128 | 0.29 | 0.17 | 0.20 | - | 0.07 | - |
|   | 192 | 0.17 | 0.22 | 0.18 | - | 0.08 | - |
|   | 256 | 0.19 | 0.19 | 0.21 | - | 0.08 | - |
| D | 128 | 0.19 | 0.20 | 0.15 | 0.27 | 0.18 | 0.17 |
|   | 192 | 0.22 | 0.18 | 0.15 | 0.23 | 0.19 | 0.22 |
|   | 256 | 0.18 | 0.23 | 0.24 | 0.17 | 0.21 | 0.17 |
| E | 128 | 0.23 | 0.24 | 0.40 | 0.26 | 0.21 | 0.19 |
|   | 192 | 0.19 | 0.32 | 0.27 | 0.31 | 0.25 | 0.31 |
|   | 256 | 0.21 | 0.33 | 0.26 | 0.29 | 0.24 | 0.27 |

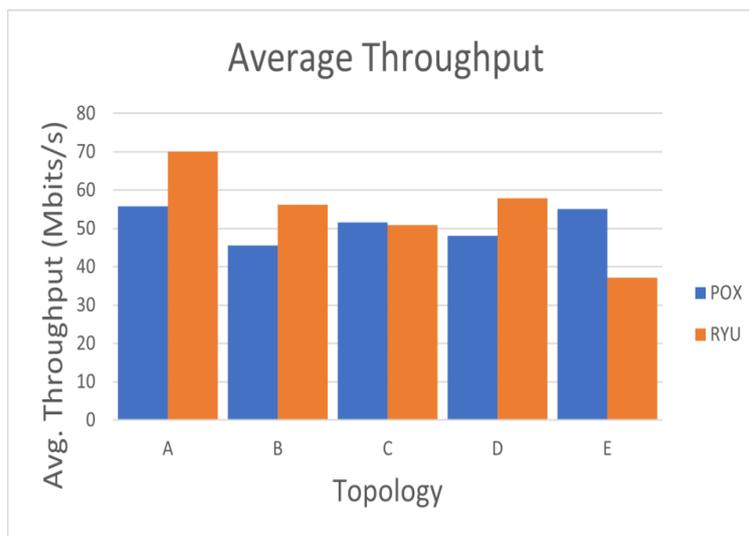

Figure 3: Throughput Comparison

Table 4: Average Delay Between Hosts (ms) – RYU

| Top. | Pkt. Size | H1to H2 | H2to H3 | H4to H5 | H6to H8 | H2to H6 | H1to H8 |
|---|---|---|---|---|---|---|---|
| A | 128 | 0.09 | - | - | - | - | - |
|   | 192 | 0.08 | - | - | - | - | - |
|   | 256 | 0.10 | - | - | - | - | - |
| B | 128 | 0.12 | 0.15 | - | - | - | - |
|   | 192 | 0.13 | 0.22 | - | - | - | - |
| C | 256 | 0.15 | 0.15 | - | - | - | - |
|   | 128 | 0.12 | 0.11 | 0.15 | - | 0.17 | - |
|   | 192 | 0.10 | 0.11 | 0.13 | - | 0.13 | - |
| D | 256 | 0.11 | 0.15 | 0.18 | - | 0.14 | - |
|   | 128 | 0.08 | 0.12 | 0.15 | 0.12 | 0.18 | 0.13 |
| E | 192 | 0.11 | 0.12 | 0.16 | 0.11 | 0.13 | 0.17 |
|   | 256 | 0.17 | 0.12 | 0.11 | 0.14 | 0.12 | 0.15 |
|   | 128 | 0.11 | 0.11 | 0.11 | 0.14 | 0.14 | 0.13 |
|   | 192 | 0.11 | 0.09 | 0.14 | 0.13 | 0.14 | 0.12 |
|   | 256 | 0.14 | 0.16 | 0.14 | 0.15 | 0.11 | 0.18 |

The overall average delay between two hosts for different packet sizes is shown in Fig.5. Values depicted in Fig.5 were calculated by averaging delays for different pairs of hists.

The data demonstrates notable differences in average delays across topologies (A to E) and varying traffic loads (128, 192, and 256 Mbps) for the POX and RYU controllers. Overall, RYU consistently achieves lower delays compared to POX, indicating superior performance in handling network traffic. For both controllers, increasing traffic loads generally results in higher delays, a trend expected due to the greater strain on network resources.

At a traffic load of 128 Mbps, RYU significantly outperforms POX across all topologies. The largest discrepancy is observed in Topology E, where RYU records a delay of 0.12387 seconds compared to POX's 0.25373 seconds, more than twice the delay. Even in Topology B, where the performance gap is smallest, RYU (0.13185 seconds) still surpasses POX (0.1704 seconds). Similar trends are evident at 192 Mbps, where RYU demonstrates significant improvements over POX, particularly in Topology E (0.12125 seconds vs. 0.2762 seconds). However, in Topology B, the difference narrows slightly, with RYU recording a delay of 0.1791 seconds compared to POX's 0.1207 seconds, indicating some limitations of RYU under certain traffic conditions.

Table 5: Average Jitter Between Hosts(ms) –POX

| Top. | Pkt. Size | H1to H2 | H2to H3 | H4to H5 | H6to H8 | H2to H6 | H1to H8 |
|---|---|---|---|---|---|---|---|
| A | 128 | 0.08 | - | - | - | - | - |
|   | 192 | 0.10 | - | - | - | - | - |
|   | 256 | 0.21 | - | - | - | - | - |
| B | 128 | 0.09 | 0.15 | - | - | - | - |
|   | 192 | 0.05 | 0.07 | - | - | - | - |
| C | 256 | 0.02 | 0.12 | - | - | - | - |
|   | 128 | 0.15 | 0.11 | 0.16 | - | 0.01 | - |
|   | 192 | 0.14 | 0.12 | 0.11 | - | 0.01 | - |
| D | 256 | 0.09 | 0.13 | 0.15 | - | 0.01 | - |
|   | 128 | 0.13 | 0.11 | 0.05 | 0.12 | 0.07 | 0.12 |
| E | 192 | 0.20 | 0.10 | 0.05 | 0.13 | 0.09 | 0.14 |
|   | 256 | 0.06 | 0.18 | 0.20 | 0.09 | 0.13 | 0.10 |
|   | 128 | 0.09 | 0.18 | 0.23 | 0.27 | 0.10 | 0.12 |
|   | 192 | 0.18 | 0.32 | 0.17 | 0.20 | 0.20 | 0.31 |
|   | 256 | 0.07 | 0.23 | 0.14 | 0.14 | 0.11 | 0.12 |

Table 6: Average Jitter Between Hosts (ms)–RYU

| Top. | Pkt. Size | H1to H2 | H2to H3 | H4to H5 | H6to H8 | H2to H6 | H1to H8 |
|---|---|---|---|---|---|---|---|
| A | 128 | 0.04 | - | - | - | - | - |
|   | 192 | 0.02 | - | - | - | - | - |
|   | 256 | 0.03 | - | - | - | - | - |
| B | 128 | 0.06 | 0.08 | - | - | - | - |
|   | 192 | 0.09 | 0.01 | - | - | - | - |
| C | 256 | 0.12 | 0.09 | - | - | - | - |
|   | 128 | 0.09 | 0.06 | 0.12 | - | 0.08 | - |
|   | 192 | 0.03 | 0.05 | 0.09 | - | 0.06 | - |
| D | 256 | 0.09 | 0.08 | 0.14 | - | 0.08 | - |
|   | 128 | 0.02 | 0.06 | 0.06 | 0.04 | 0.21 | 0.05 |
| E | 192 | 0.04 | 0.05 | 0.10 | 0.09 | 0.03 | 0.13 |
|   | 256 | 0.08 | 0.04 | 0.03 | 0.04 | 0.04 | 0.12 |
|   | 128 | 0.06 | 0.06 | 0.08 | 0.11 | 0.13 | 0.10 |
|   | 192 | 0.03 | 0.02 | 0.10 | 0.08 | 0.11 | 0.10 |
|   | 256 | 0.11 | 0.15 | 0.06 | 0.12 | 0.07 | 0.12 |

At the highest traffic load of 256 Mbps, both controllers experience increased delays. RYU maintains its advantage in most topologies, particularly in Topology A, where it records a delay of 0.0995 seconds compared to POX's 0.2275 seconds. Interestingly, in Topology B, POX achieves a slightly lower delay of 0.1181 seconds, outperforming RYU's 0.1482 seconds—a rare instance where POX proves more efficient under higher traffic conditions.

Examining performance across topologies, RYU consistently demonstrates lower delays in Topologies A, C, D, and E across all traffic loads. In Topology A, RYU achieves delays below 0.1 seconds under all traffic conditions, while POX records significantly higher delays, particularly at 256 Mbps (0.2275 seconds). In Topology C, RYU's advantage is particularly evident at 192 Mbps, with a delay of 0.11775 seconds compared to POX's 0.16155 seconds. In Topology D, both controllers exhibit similar trends, but RYU maintains lower delays at all traffic loads, particularly at 192 Mbps, where it records 0.13603 seconds compared to POX's 0.1975 seconds. Topology E highlights RYU's dominance, particularly under higher traffic loads, where its delay at 256 Mbps (0.14688 seconds) is significantly lower than POX's (0.26917 seconds).

In summary, RYU demonstrates superior delay performance across most topologies and traffic loads, particularly in scenarios with heavier traffic and complex network configurations, such as Topology E. POX, however, performs competitively in specific scenarios, particularly in Topology B under higher traffic loads, where it achieves the lowest delay at 256 Mbps. These results suggest that the choice of controller should be informed by the specific network topology and expected traffic load, with RYU being a more suitable option for scenarios requiring low delays and high traffic efficiency.

### 4.3. Jitter Comparison

Jitter was measured between the hosts H1-H2, H2-H3, H4-H5, H6-H8, H2-H6, and H1-H8 (refer to 2 covering all applicable topologies. The average jitter between two hosts is shown in Fig.5.

The analysis of average jitter across topologies A to E under different packet sizes (128, 192, and 256 Bytes) for POX and RYU controllers highlights significant differences in performance. Jitter, a measure of packet delay variation, directly impacts the quality of service in networks. Across all scenarios, RYU demonstrates consistently lower jitter compared to POX, indicating better network stability and reliability.

For 128 Byte packets, RYU significantly outperforms POX in all topologies. For instance, in Topology A, RYU records a jitter of 0.0383 seconds, which is less than half of POX's 0.0833 seconds. The largest performance gap is observed in Topology B, where RYU achieves a jitter of 0.0704 seconds compared to POX's 0.11835 seconds. Even in Topology E, where the values are closer, RYU (0.08995 seconds) maintains a clear advantage over POX (0.10128 seconds). This trend continues for 192 Byte packets, where RYU exhibits exceptionally low jitter, particularly in Topology A, with a value of 0.0219 seconds compared to POX's 0.0951 seconds. In Topology C, RYU achieves a jitter of 0.05928 seconds, significantly better than POX's 0.0943 seconds. Similarly, in Topology B, although the gap narrows, RYU still outperforms POX with jitter values of 0.04675 seconds and 0.05875 seconds, respectively.

At the highest packet size of 256 Bytes, jitter increases for both controllers, reflecting the challenges of maintaining network stability under heavy traffic. However, RYU continues to deliver better performance in most scenarios. In Topology A, RYU records a jitter of 0.0342 seconds, a stark contrast to POX's 0.2116 seconds, highlighting RYU's efficiency even under significant traffic loads. Similarly, in Topology D, RYU achieves 0.06018 seconds compared to POX's 0.12545 seconds. Interestingly, in Topology B, POX performs competitively, achieving a jitter of 0.06595 seconds compared to RYU's 0.1066 seconds. This is one of the few instances where POX outperforms RYU, suggesting its suitability in specific topologies under heavy traffic.

When analysed by topology, RYU demonstrates remarkable stability in Topologies A, C, and D. In Topology A, RYU consistently achieves jitter values below 0.04 seconds across all traffic loads, while POX's jitter increases significantly, particularly for 256 Byte packets (0.2116 seconds). In Topology C, RYU's jitter remains significantly lower than POX's at all traffic loads, demonstrating its superiority in this topology. Similarly, in Topology D, RYU maintains low jitter values, particularly for 256 Byte packets, where it records 0.06018 seconds compared to POX's 0.12545 seconds. Topology E also highlights RYU's advantage, though the gap narrows slightly at higher traffic loads, with RYU recording 0.10368 seconds and POX 0.12545 seconds for 256 Byte packets.

Overall, RYU exhibits superior performance in minimising jitter across most topologies and traffic loads, with particular strengths in Topologies A, C, and D. POX, while generally less stable, demonstrates competitive performance in Topology B under higher traffic conditions. These findings suggest that RYU is better suited for scenarios requiring low jitter and consistent network performance, while POX may have advantages in specific topologies with unique traffic patterns.

## 5. CONCLUSION

This study systematically evaluated the Quality of Service (QoS) performance of the POX and RYU SDN controllers across five distinct network topologies using Mininet. Given the increasing reliance on SDN for efficient network management, understanding the QoS capabilities of different controllers is crucial for optimising performance across diverse network environments.
Through a structured three-stage methodology—comprising simulation environment design, controller rule implementation, and performance evaluation—this research provided a comprehensive comparison of POX and RYU in terms of throughput, delay (round-trip time), and jitter. These metrics, which directly influence network efficiency and user experience, were carefully analysed across various topologies, ranging from a network with a single Layer 2 switch to more complex network structures.

The results demonstrated that RYU consistently outperformed POX in terms of throughput, particularly in Topologies A, B, and D, suggesting its superior capability in handling high-bandwidth applications. However, POX exhibited better performance in Topology E, indicating that its lightweight architecture can be advantageous in specific network conditions. In terms of delay, RYU achieved consistently lower latency across most scenarios, reinforcing its suitability for real-time applications such as VoIP and video streaming. Additionally, jitter analysis confirmed that RYU maintained more stable packet transmission, a critical factor for time-sensitive communication.

These findings align with existing research, which highlights RYU's advanced traffic management capabilities and scalability compared to POX. However, POX remains a viable option for specific use cases, particularly where simplicity and ease of implementation are prioritised. The study underscores the importance of selecting an SDN controller based on the specific QoS requirements of the network, ensuring optimal performance for different applications.

Future research could extend this work by incorporating additional SDN controllers, assessing performance under varying traffic loads, and exploring scalability in large-scale deployments. A deeper investigation into adaptive QoS mechanisms within SDN environments could further enhance network performance and reliability. By refining our understanding of SDN controller capabilities, this research contributes to the ongoing development of more efficient, flexible, and resilient network infrastructures.

## CONFLICT OF INTEREST

The authors declare no conflict of interest.

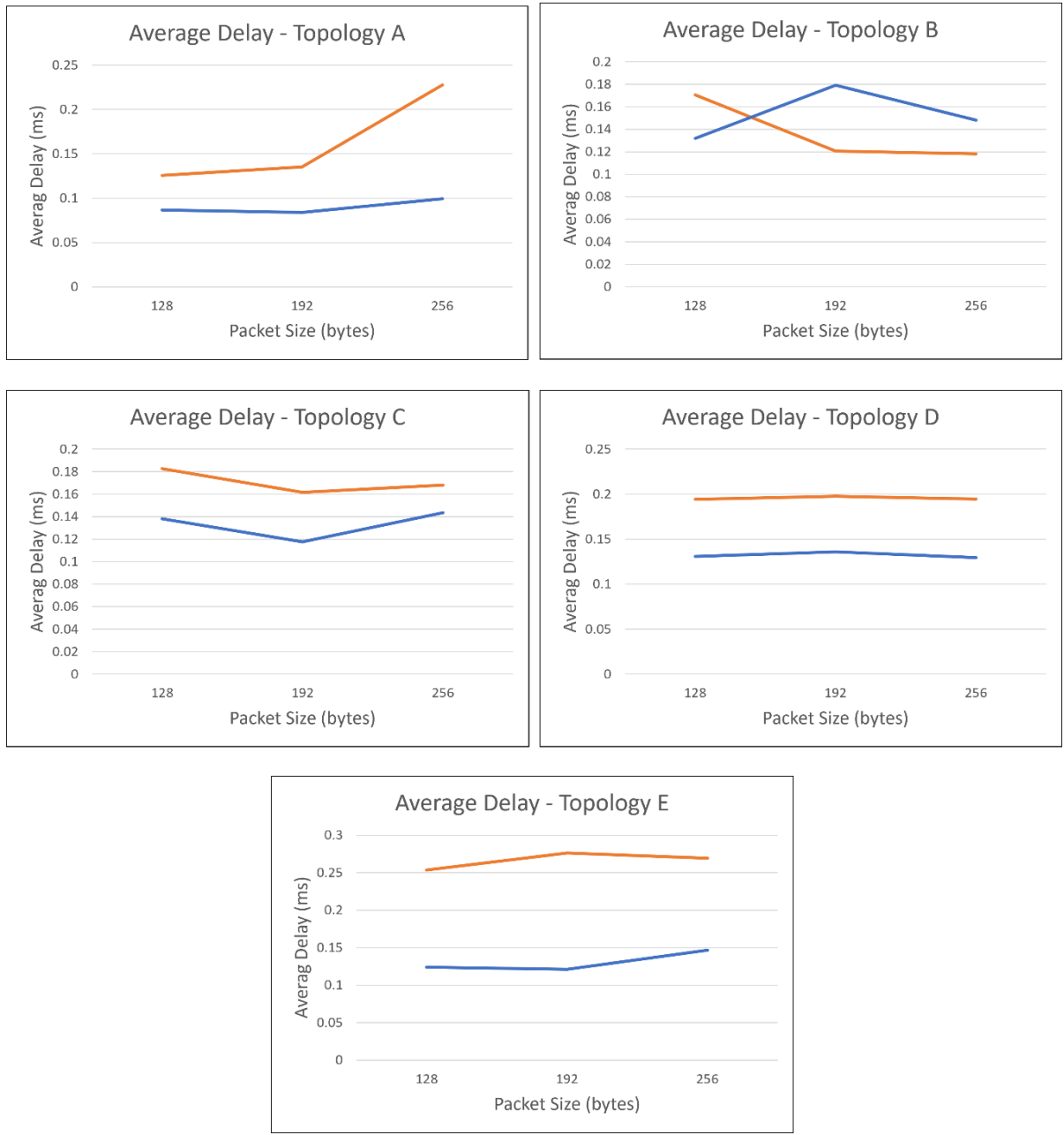

Figure 4: Delay Comparison

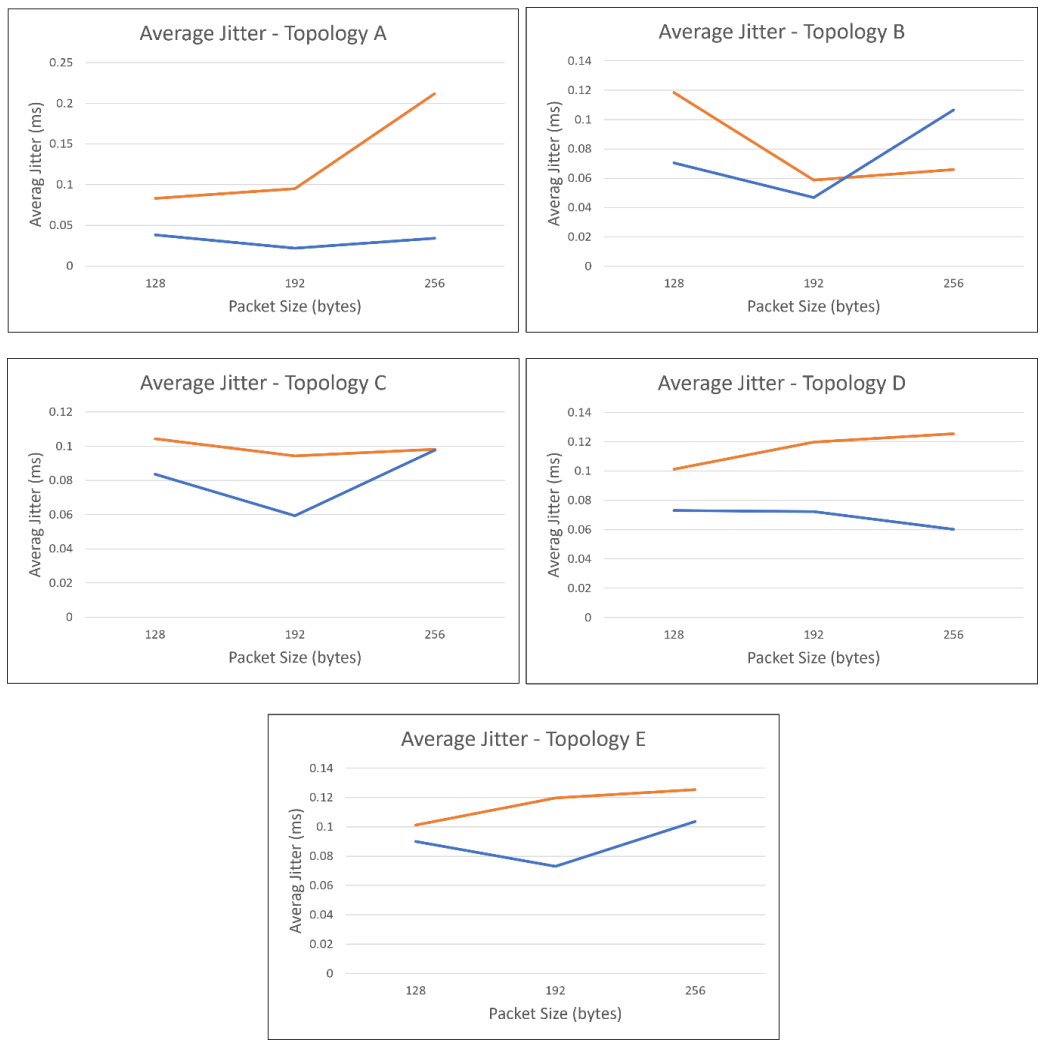

Figure5: Jitter Comparison